\documentclass[onecolumn,aps,pra,reprint,superscriptaddress,longbibliography]{revtex4-1}

\usepackage{amsmath}
\usepackage{graphicx}
\usepackage[lofdepth,lotdepth, caption=false]{subfig}
\usepackage{verbatim}
\usepackage{color}
\usepackage{dcolumn}
\usepackage{bm}
\usepackage{miller}
\usepackage{color}

\usepackage{xr-hyper}
\usepackage{hyperref}

\usepackage[english]{babel}

\newcommand{\beginsupplement}{%
        \setcounter{table}{0}
        \renewcommand{\thetable}{S\arabic{table}}%
        \setcounter{figure}{0}
        \renewcommand{\thefigure}{S\arabic{figure}}%
     }

\begin{document}

\title{Helimagnetism in CrBr$_2$ and CrIBr}

\author{John A.~Schneeloch}
\affiliation{Department of Physics, University of Virginia, Charlottesville,
Virginia 22904, USA}

\author{Adam A.\ Aczel}
\affiliation{Neutron Scattering Division, Oak Ridge National Laboratory, Oak Ridge, Tennessee 37831, USA}

\author{Despina Louca}
\thanks{Corresponding author}
\email{louca@virginia.edu}
\affiliation{Department of Physics, University of Virginia, Charlottesville,
Virginia 22904, USA}

\begin{abstract}    
In CrX$_2$ (X=Br, I), a Jahn-Teller effect distorts the octahedral configuration of the anions about Cr, resulting in a ``ribbon chain'' structure.
We previously observed helimagnetism in CrI$_2$ propagating along the ribbon chains, with an 80-90$^{\circ}$ rotation per Cr ion. 
Via neutron scattering, we report that CrBr$_2$ and the solid solution CrIBr are also helimagnetic, with N\'{e}el temperatures of 17 and 12 K, respectively, and signs of intermediate magnetic transitions in both compounds. The helical angle between spins on consecutive Cr ions along the ribbon chains increases with increasing Br substitution, from  89.7$^{\circ}$ (CrI$_2$) to 116$^{\circ}$ (CrIBr) to 147$^{\circ}$ (CrBr$_2$), possibly as a result of decreasing longer-range intrachain spin interactions with the increased electron localization of the lighter anions. 
\end{abstract}

\maketitle

\section{Introduction}
Van der Waals (vdW) layered magnetic materials have received much attention recently due to interest in the 2-dimensional materials that make up the structure of these compounds, but many such materials have, nevertheless, remained unexplored. These materials may have interesting properties, such as the helimagnetic ordering that we recently identified in CrI$_2$ \cite{schneelochHelimagnetismCandidateFerroelectric2024}. Jahn-Teller distortion in CrX$_2$ (X=Cl, Br, I) causes the CrX$_6$ octahedra to distort into two long and four short Cr-X bonds, giving rise to ``ribbon chains'' of edge-sharing octahedra along a crystallographic axis. 
Ribbon chains are also found in a number of Cu$^{2+}$ compounds, and the magnetism of these materials have received much study, including in CuCl$_2$, CuBr$_2$, and in copper oxides such as LiCu$_2$O$_2$ and LiCuVO$_4$ \cite{xiangDensityFunctionalCharacterizationMultiferroicity2007} which have a spin-spiral magnetism that is strikingly similar to CrI$_2$ \cite{banksMagneticOrderingFrustrated2009, zhaoCuBr2NewMultiferroic2012}. 
Notably, CuCl$_2$ \cite{sekiCupricChlorideTextCuCl2010} and CuBr$_2$ \cite{zhaoCuBr2NewMultiferroic2012} exhibit multiferroic behavior, in which an electric polarization follows the onset of cycloidal magnetic order \cite{tokuraMultiferroicsSpinOrigin2014}. 
The search for a monolayer multiferroic of this type is an active area of research, with NiI$_2$ as a possible candidate \cite{songEvidenceSinglelayerVan2022,jiangDilemmaOpticalIdentification2023}. 
Monolayers of both CrI$_2$ \cite{liSinglelayerCrI3Grown2020,caiMolecularBeamEpitaxy2021,liTwoDimensionalMagneticSemiconducting2023} and CrBr$_2$ \cite{karjasiltaMolecularBeamEpitaxy2023,kezilebiekeElectronicMagneticCharacterization2021} have been synthesized, making these materials additional candidates for monolayer multiferroicity. 
More generally, understanding the magnetic properties of CrX$_2$ should improve our understanding of related magnetic materials. 
CrBr$_2$, to our knowledge, has not been studied and its magnetic properties are unknown, making it one of the last of the $3d$ transition metal halides to be thus investigated \cite{mcguireReview}.

\begin{figure}[t]
\begin{center}
\includegraphics[width=8.6cm]
{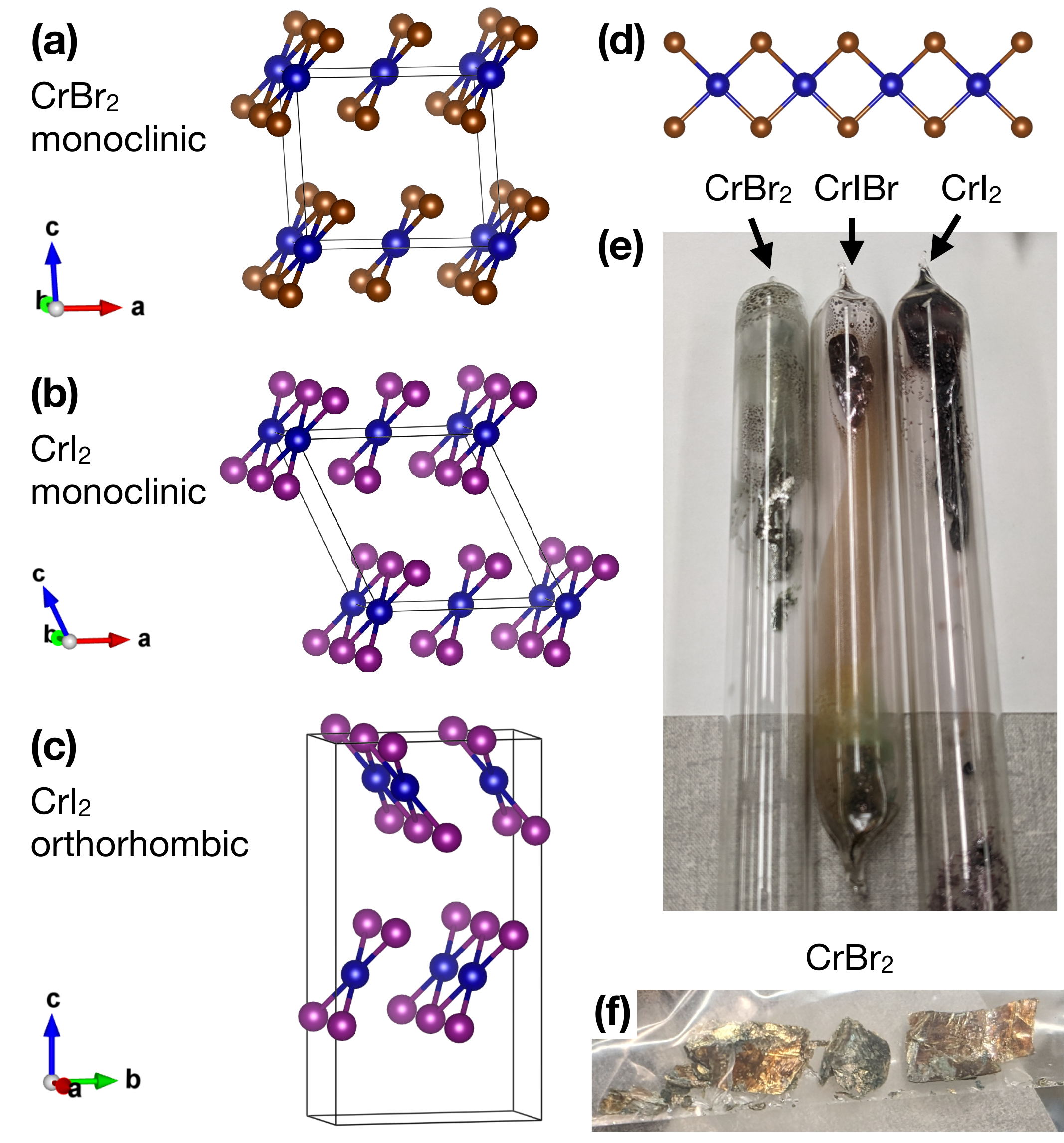}
\end{center}
\caption{(a-c) Crystal structures of monoclinic CrBr$_2$ (a), monoclinic CrI$_2$ (b), and orthorhombic CrI$_2$ (c). Cr$^{2+}$ ions are blue, I$^{-}$ are magenta, and Br$^{-}$ are brown. These structures form ribbon chains along the \emph{b}, \emph{b}, and \emph{a} axes, respectively, shown in extended form in (d). Structures plotted using \textsc{VESTA} \cite{mommaVESTAThreedimensionalVisualization2011}. (e) Photo of three ampoules (with inner diameters of 20 mm) containing crystals of CrBr$_2$, CrIBr, and CrI$_2$. A waxy CuX$_2$ condensation can be seen on the inner surface of each ampoule. (f) Photo of CrBr$_2$ crystals inside of a plastic bag, with in-plane widths of about 1 cm.}
\label{fig:schematic}
\end{figure}

In CrI$_{x}$Br$_{2-x}$, the ribbon chains form layers which can be stacked in three ways, as depicted in Fig.\ \ref{fig:schematic}(a-c). 
(Ribbon chains are also present in CrCl$_2$, but they do not form the layers seen in CrI$_{x}$Br$_{2-x}$ \cite{tracyCrCl2Paper,winkelmannStructuralMagneticCharacterization1997}.)
There are two monoclinic structures with the layers oriented in the same direction, corresponding to CrBr$_2$ with $\beta \approx 94^{\circ}$ \cite{tracyCrBr2Paper} (Fig.\ \ref{fig:schematic}(a)) and CrI$_2$ with $\beta = 115.5^{\circ}$\cite{tracyCrI2MonoPaper} (Fig.\ \ref{fig:schematic}(b).) The monoclinic CrBr$_2$ (mono-CrBr$_2$) and CrI$_2$ (mono-CrI$_2$) structures can be thought of as distorted versions of the CdI$_2$-type and CdCl$_2$-type structures \cite{mcguireReview}, respectively, or as different arrangements of close-packed atoms \cite{tracyCrBr2Paper}. The mono-CrI$_2$ layer stacking has an additional in-plane translation of roughly $-\frac{1}{3}$ along the $a$-axis relative to mono-CrBr$_2$, for which the layers are nearly perpendicularly stacked. 
The orthorhombic CrI$_2$ (ortho-CrI$_2$) structure, meanwhile, has layers with alternating orientation \cite{besrestStructureCristallineIodure1973} (Fig.\ \ref{fig:schematic}(c)), breaking inversion symmetry in a manner reminiscent of the $T_d$ phase of MoTe$_2$ \cite{clarkeLowtemperatureStructuralPhase1978}, WTe$_2$ \cite{brownCrystalStructuresWTe21966}, or $\alpha$-ZrI$_2$ 
\cite{corbettSecondInfinitechainForm1982,guthrieSynthesisStructureInfinitechain1981}. 
In our study of CrI$_2$ \cite{schneelochHelimagnetismCandidateFerroelectric2024}, we have typically observed the orthorhombic structure, though the crystal we measured with single-crystal x-ray diffraction (SCXRD) had a minority mono-CrI$_2$ phase that was aligned with the main ortho-CrI$_2$ phase. Disordered orthorhombic-type stacking was present in our powder CrI$_2$ sample. 

The magnetism of ortho-CrI$_2$ consists of a helical order below $T_N=17$ K with an 89.7$^{\circ}$ helical angle (spin rotation per Cr ion along the ribbon chain) at 8 K, corresponding to a propagation wavevector  of $(0.2492,0,0)$ \cite{schneelochHelimagnetismCandidateFerroelectric2024}. 
CuCl$_2$ and CuBr$_2$, which are isostructural to mono-CrI$_2$, exhibit a similar spin spiral ordering, with wavevectors of (1,0.2257,0.5) for CuCl$_2$ \cite{banksMagneticOrderingFrustrated2009} and (1,0.2350,0.5) for CuBr$_2$ \cite{zhaoCuBr2NewMultiferroic2012}, corresponding to spin rotation angles along consecutive Cu$^{2+}$ ions along the ribbon chains of 81.3$^{\circ}$ and 84.6$^{\circ}$, respectively. The spin-spiral plane is different, though, being cycloidal for the copper dihalides rather than helical for CrI$_2$. 
The primary cause of the spin spiral ordering in these compounds is speculated to be a frustration between a strongly antiferromagnetic (AFM) next-nearest-neighbor (NNN) intrachain interaction, and a sufficiently weak nearest-neighbor (NN) interaction. 
Such a situation can be modeled by a linear $J_1$-$J_2$ Heisenberg model where $J_1$ and $J_2$ are the NN and NNN exchange couplings, respectively, between spins on a ribbon chain; if $J_2 < 0$ (is AFM) and $|J_1| < 4|J_2|$, spin spiral ordering results   \cite{blundellMagnetismCondensedMatter2001}.  Density functional theory (DFT) calculations have supported this interpretation for CuCl$_2$  \cite{banksMagneticOrderingFrustrated2009}, CuBr$_2$ \cite{leeInvestigationSpinExchange2012} and ortho-CrI$_2$ \cite{schneelochHelimagnetismCandidateFerroelectric2024}. 

These materials differ in the degree of interchain magnetic coupling. The copper dihalides exhibit signs of quasi-one-dimensionality, with a broad hump in the magnetic susceptibility characteristic of low-dimensional systems \cite{banksMagneticOrderingFrustrated2009,zhaoCuBr2NewMultiferroic2012}. The susceptibility of CrCl$_2$ (which has collinear AFM order along the ribbon chains \cite{cableNeutronDiffractionStudies1960}) also has a broad hump and can be modeled well by an AFM spin chain \cite{hagiwaraMagneticPropertiesAnhydrous1995}, and inelastic neutron scattering data suggest that the interchain coupling is about $\sim$5\% that of the intrachain NN coupling \cite{stoneQuasionedimensionalSpinWaves2013}. In contrast, the susceptibility of CrI$_2$ exhibits Curie-Weiss behavior, with only a very subtle deviation indicating the onset of magnetic ordering \cite{schneelochHelimagnetismCandidateFerroelectric2024}. 
Additionally, DFT studies suggest a substantial interlayer magnetic coupling for ortho-CrI$_2$ of a similar magnitude to the intrachain NN and NNN exchange constants  \cite{schneelochHelimagnetismCandidateFerroelectric2024}. 
An increase in interlayer coupling is expected as one replaces lighter halogens for heavier ones, as seen from nuclear magnetic resonance measurements on CrCl$_3$ \cite{narathSpinWaveAnalysisSublattice1965}, CrBr$_3$ \cite{davisSpinWaveRenormalizationApplied1964}, and CrI$_3$ \cite{narathZeroField53MathrmCr1965}. 
As for CrBr$_2$, though there has been an absence of experimental evidence, one expects the magnetism of this compound to resemble that of CrI$_2$. 

To explore the magnetism of CrBr$_2$ and the intermediate composition CrIBr, we conducted single-crystal elastic neutron scattering measurements. In both compounds, helimagnetic order propagates along the ribbon chains, very similar to that seen in CrI$_2$, but with a substantial change in helical angle. In CrBr$_2$, helimagnetic order appears below 17 K with a propagation vector of (1, 0.40845(15), 0.5) at 3.7 K, corresponding to a helical angle of 147.0$^{\circ}$. 
In CrIBr, helimagnetic order appears below 12 K with a propagation vector of ($\delta$, 0, 0) with $\delta=0.32216(5)$ at 3.8 K, corresponding to a helical angle of 116.0$^{\circ}$, nearly the average of those of the end members CrI$_2$ (89.7$^{\circ}$) and CrBr$_2$.

\section{Experimental Details}
Single crystals of CrBr$_2$ were grown by a direct reaction of a stoichiometric amount of the elements. 
Cr powder (-200 mesh size) was added to an open-ended glass tube. After making the constriction for the ampoule, bromine was weighed in a beaker and poured down the open end of the tube. (Given the high vapor pressure of bromine, it was difficult to control the starting ratio of the elements to better than a few percent.) 
Liquid nitrogen was used to condense the vapor and freeze the bromine, followed by purging with argon and sealing the ampoule in vacuum. The CrBr$_2$ crystals were grown by slowly cooling the ampoule from 900 $^{\circ}$C to 750 $^{\circ}$C over several hours, then quenching in water. (The reported melting point of CrBr$_2$ is 842 $^{\circ}$C \cite{perryHandbookInorganicCompounds2011}.) Crystals of CrIBr were synthesized by  sealing stoichiometric amounts of Cr powder and iodine monobromide in an ampoule, using liquid nitrogen to condense the vapor prior to sealing. The ampoule was slowly cooled from 900 $^{\circ}$C to 750 $^{\circ}$C, then quenched in water. 
The CrI$_{2-x}$Br$_x$ compounds are hygroscopic and may decompose into a green liquid within minutes of exposure to non-dry air, though samples can be stored in dry air for an extended period of time. The color of CrBr$_2$ can range from a pale green to an amber brown, depending on the angle (Fig.\ \ref{fig:schematic}(e,f)), while CrIBr is red and CrI$_2$ is a darker red (Fig.\ \ref{fig:schematic}(e).) 

Magnetization measurements were performed in a Quantum Design Physical Property Measurement System equipped with a Vibrating Sample Magnetometer. Data were taken continuously on cooling or warming at a rate of 1 K/min. 

Single-crystal neutron scattering measurements were performed on the triple-axis spectrometer VERITAS at the High Flux Isotope Reactor at Oak Ridge National Laboratory. The incident neutron energy was $E_i=14.45$ meV, and the collimations were set to 40'-40'-S-40'-80'. The samples were a 34 mg CrBr$_2$ crystal and an 80 mg CrIBr crystal. Samples were coated with the fluorine-based glue CYTOP to protect them from humidity and to secure them to aluminum plates prior to measurement. A top-loading closed-cycle refrigerator was used.

Single crystal x-ray diffraction (SCXRD) was carried out at various temperatures on  CrBr$_2$ and CrIBr crystals, with data analyzed using the software \textsc{APEX5} by Bruker. 

\section{Results}

\subsection{Single-crystal x-ray diffraction}
\label{sec:SCXRD}

\begin{figure*}[t]
\begin{center}
\includegraphics[width=18cm]
{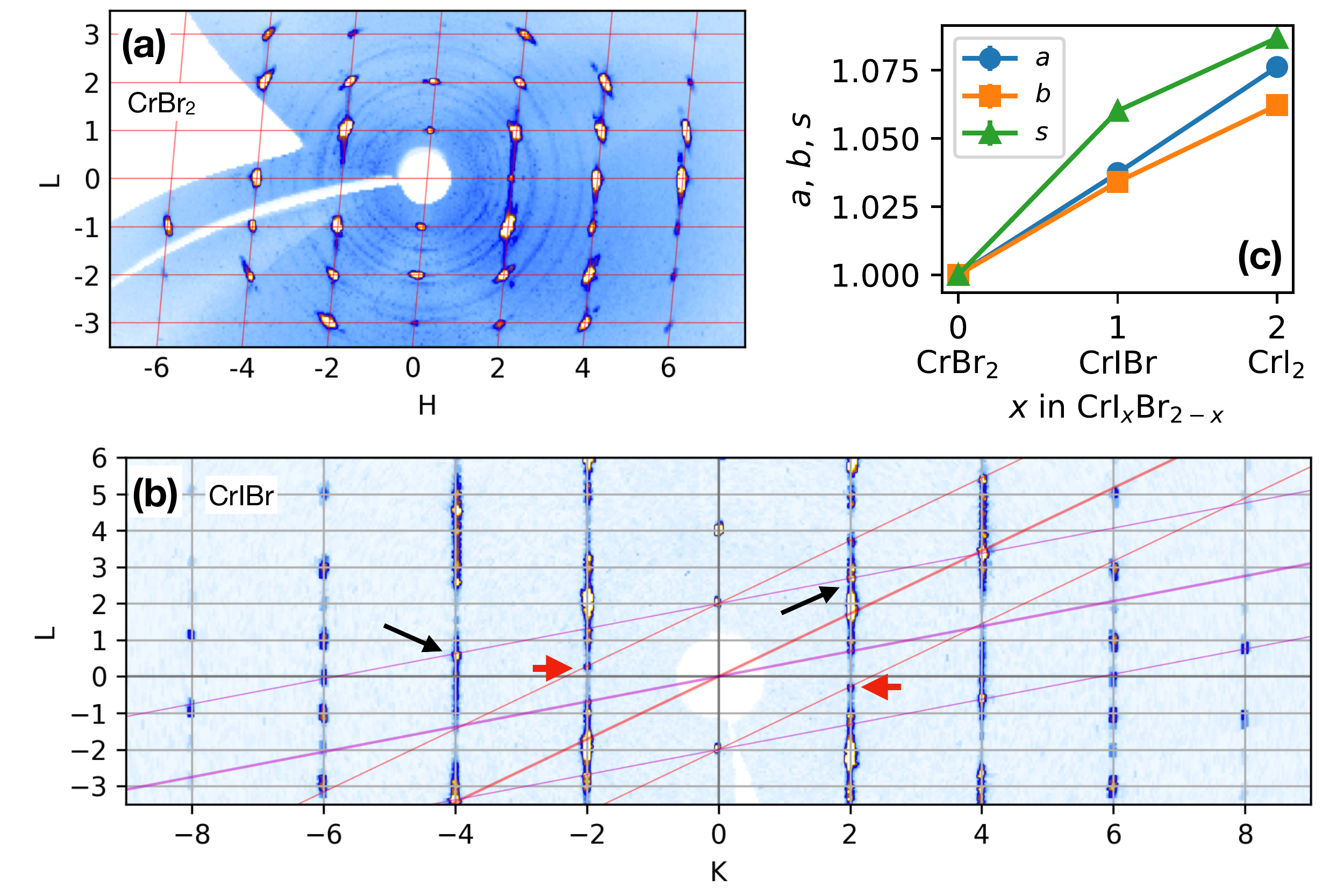}
\end{center}
\caption{(a) SCXRD intensity in the $H0L$ plane for CrBr$_2$ at 300 K. (b) SCXRD intensity in the $0KL$ plane for CrIBr at 300 K. The gray grid lines and the labels on the axes correspond to the orthorhombic phase. 
Peaks from two secondary monoclinic phases are indicated by arrows: a mono-CrI$_2$-type phase (short red arrows), and a $1T^{\prime}$-type phase (long black arrows.) These phases all share the vertical axis, i.e., [001]. For the mono-CrI$_2$-type phase, the red lines denote $(H0L)_{\textrm{mono}}$ for $L=0, \pm1$, where $H_{\textrm{mono}} = K$. For the $1T^{\prime}$-type phase, the magenta lines denote $(H0L)_{1T^{\prime}}$ for $L=0, \pm2$, where $H_{1T^{\prime}} = K$. 
(c) In-plane lattice parameters and interlayer spacing as a function of nominal composition, as determined by room-temperature SCXRD measurements. The parameters $a$ and $b$ represents the orthorhombic lattice parameters or the monoclinic equivalent (i.e., $a$ is  $a_{\textrm{ortho}}$ or $ b_{\textrm{mono}}$, representing the ribbon-chain axis; and the other in-plane direction, $b$, is $b_{\textrm{ortho}}$ or $a_{\textrm{mono}}$.) The interlayer spacing $s$ is $c_{\textrm{ortho}}/2$ or $c_{\textrm{mono}} \cos(\beta)$. All parameters are normalized to their CrBr$_2$ values. CrI$_2$ values are from Ref.\ \cite{schneelochHelimagnetismCandidateFerroelectric2024}. Error bars are smaller than markers.}
\label{fig:SCXRD}
\end{figure*}

\begin{table}[t]
\caption{Lattice parameters from SCXRD refinement for CrBr$_2$.}
\label{tab:SCXRD1}
\begin{tabular}{lllll}
\hline
 & 100 K & 200 K & 300 K & 400 K  \\
\hline
$a$ (\AA) & 7.099(7) & 7.116(9) & 7.129(10) & 7.157(7) \\
$b$ (\AA) & 3.649(4) & 3.646(6) & 3.642(6) & 3.639(4) \\
$c$ (\AA) & 6.242(9) & 6.256(12) & 6.258(13) & 6.256(8) \\
$\beta$ ($^{\circ}$) & 94.20(4) & 94.29(4) & 94.50(5) & 94.59(4) \\
\hline
\end{tabular}
\end{table}

In Figures \ref{fig:SCXRD}(a) and \ref{fig:SCXRD}(b) we show SCXRD intensity at 300 K for CrBr$_2$ and CrIBr, respectively. For both materials, the reciprocal-space plane shown is the one perpendicular to the ribbon chains, but due to their different unit cells (as shown in Fig.\ \ref{fig:schematic}(a-c)) these planes are labeled differently. Specifically, in CrBr$_2$ (Fig.\ \ref{fig:schematic}(a)), the ribbon chains run along the $b$-axis and there is one layer per unit cell, while for CrIBr, which largely has the ortho-CrI$_2$ structure, the ribbon chains run along the $a$-axis and there are two layers per unit cell. Only one mono-CrBr$_2$ twin is present in Fig.\ \ref{fig:SCXRD}(a), with some diffuse scattering that is likely due to mono-CrI$_2$-type stacking defects (roughly a few percent according to our calculations.) The refined lattice parameters for CrBr$_2$ are shown in Table \ref{tab:SCXRD1}, which disagree somewhat with the previous reported values of $a=7.114$ \AA, $b=3.649$ \AA, $c=6.217$ \AA, and $\beta=93.88^{\circ}$ \cite{tracyCrBr2Paper}. 
Most of the parameters show the expected thermal expansion behavior; there might be a negative thermal expansion along the ribbon-chain axis (the $b$-axis), but the uncertainties are too large to say for certain. 

The CrIBr data in Fig.\ \ref{fig:SCXRD}(b) largely corresponds to an ortho-CrI$_2$-type phase with room-temperature lattice parameters of $a=3.778(3)$ \AA, $b=7.372(6)$ \AA, and $c=13.227(11)$ \AA\ as determined from SCXRD. 
The value of $a$ is as expected given a linear trend between the corresponding CrBr$_2$ and CrI$_2$ endpoints, as seen in Fig.\ \ref{fig:SCXRD}(c), suggesting that the true composition of CrIBr is very close to the nominal composition. (A similar room-temperature value was seen on a different crystal measured by neutron scattering, with $a=3.7824(19)$ \AA\ obtained via the location of the $(200)$ Bragg peak, suggesting that there was little compositional variation between the different CrIBr crystals synthesized from chromium powder and iodine monobromide.) 
The other two lattice constants, though, are greater for CrIBr than a linear trend would suggest, slightly so for $b$ and especially so for $c$ (plotted as the interlayer spacing $s=c/2$ in Fig.\ \ref{fig:SCXRD}(c).) These trends are as expected given the weaker interlayer or intralayer-interchain bonding as compared with the intrachain bonding, since the larger I$^-$ ions would have a disproportionate impact in determining how the atoms are spaced. 

The CrIBr SCXRD data (Fig.\ \ref{fig:SCXRD}) show a significant amount of diffuse scattering in the form of streaks along the $L$-direction, as well as two secondary monoclinic phases. 
As discussed previously \cite{schneelochHelimagnetismCandidateFerroelectric2024,schneelochEvolutionStructuralTransition2020}, the ortho-CrI$_2$ structure can be constructed from two symmetry-related stacking operations labeled ``A'' and ``B'', with the two ortho-CrI$_2$ twins corresponding to sequences of AAAA...\ and BBBB..., though a disordered A/B stacking sequence is also possible. 
In CrI$_2$, the A and B stacking types are related by an in-plane translation of 0.344 lattice units (l.u.) along the $b$-axis. The lack of diffuse scattering at $K=6$, and its presence at $K=2$ and 4, is consistent with a disordered A/B stacking sequence since random translations of $\pm$0.353 l.u.\ along the $b$-axis contribute phases of $e^{i 2\pi K \Delta_l}$ to each layer in the summation of the structure factor, where $\Delta_l$ is the cumulative displacement along the $b$-axis of the $l$th layer. Since 0.353 $\approx$ 1/3, $\Delta_l \approx n/3$ for some integer $n$, and thus $e^{i 2\pi K \Delta_l}$ is about unity for $K=6$ regardless of the specific stacking sequence, leading to greatly diminished diffuse scattering along $(06L)$.

There appears to be a mono-CrI$_2$-type phase (mono-CrIBr) present in the data of Fig.\ \ref{fig:SCXRD}(b), with two peaks indicated by short red arrows. The red lines indicate $(H0L)_{\textrm{mono}}$ for $L_{\text{mono}}=-1$, 0, and 1, related to the orthorhombic-phase coordinates by $L = 2 (L_{\textrm{mono}} = \epsilon H_{\textrm{mono}})$ with $\epsilon \approx 0.43$. These lines were drawn assuming the mono-CrIBr phase has the same $\beta$ angle as the mono-CrI$_2$ phase, but with the interlayer spacing and in-plane lattice parameters of the ortho-CrIBr phase. Only one twin of mono-CrIBr appears to be present, very similar to SCXRD data taken on CrI$_2$ \cite{schneelochHelimagnetismCandidateFerroelectric2024}. Presumably, the mono-CrIBr phase formed on synthesis of the crystal, since it seems unlikely that the layer direction would change on handling at low temperature, unlike the layer-translation stacking defects that may form due to cutting, etc.

An additional monoclinic secondary phase is present that appears to be consistent with a 1T$^{\prime}$-type phase, that is, a monoclinic phase formed from an ABAB...\ stacking sequence similar to that of 1T$^{\prime}$-MoTe$_2$ \cite{clarkeLowtemperatureStructuralPhase1978,schneelochEvolutionStructuralTransition2020}. In the same way that the $\beta$ angle of 1T$^{\prime}$-MoTe$_2$ can be predicted given the atomic positions of $T_d$-MoTe$_2$ \cite{schneelochEvolutionStructuralTransition2020}, we use the atomic positions of ortho-CrI$_2$ to predict the monoclinic $\beta$ angle of a $1T^{\prime}$-type CrIBr phase. 
The distance along the $b$-axis between the A and B stacking positions is about 0.353 l.u.\  \cite{schneelochHelimagnetismCandidateFerroelectric2024}; assuming that this value holds for CrIBr (though it could certainly change somewhat with composition, as $\beta$ does for Mo$_{1-x}$W$_x$Te$_2$ \cite{schneelochEvolutionStructuralTransition2020}), then the $\beta$ angle of a 1T$^{\prime}$-type monoclinic phase is given by $\beta = 90^{\circ} + \arctan(0.353$ $b_{\textrm{ortho}} / c_{\textrm{ortho}}) \approx 101^{\circ}$. On drawing the corresponding $(H0L)_{1T^{\prime}}$ lines for $L=-2$, 0, and 2 (magenta lines in Fig.\ \ref{fig:SCXRD}(b)), we see that these lines do, indeed, pass through peaks that are otherwise unexplained, such as those indicated by the black arrows. 
Unlike MoTe$_2$ \cite{clarkeLowtemperatureStructuralPhase1978} (but like $\alpha$-ZrI$_2$ \cite{corbettSecondInfinitechainForm1982}), no changes in SCXRD intensity, including layer stacking, have been observed with changes in temperature within 100 to 300 K.

\subsection{Neutron scattering}
\begin{figure}[t]
\begin{center}
\includegraphics[width=8.6cm]
{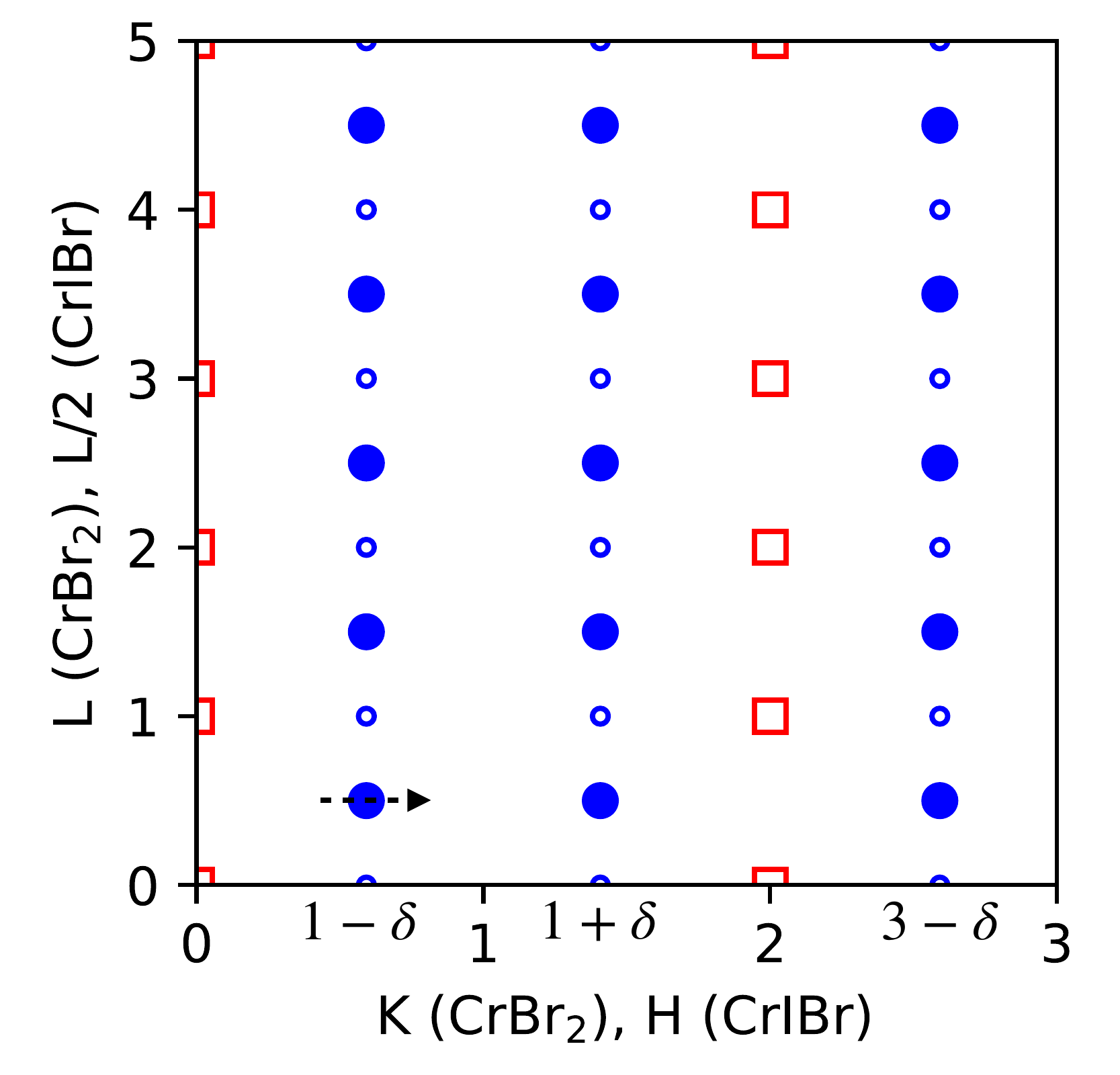}
\end{center}
\caption{Illustration of the $0KL$  scattering plane of CrBr$_2$ and the essentially equivalent $H0L$ scattering plane of CrIBr, as measured by neutrons. Red squares denote the nuclear peaks, and the blue circles show the magnetic peaks. For CrIBr, smaller additional magnetic peaks were seen, indicated by the small open circles. Most scans were taken along the $K$-direction (the $H$-direction for CrIBr), with an example shown by the dashed line.}
\label{fig:reciprocalSpaceMap}
\end{figure}

\begin{figure}[t]
\begin{center}
\includegraphics[width=8.6cm]
{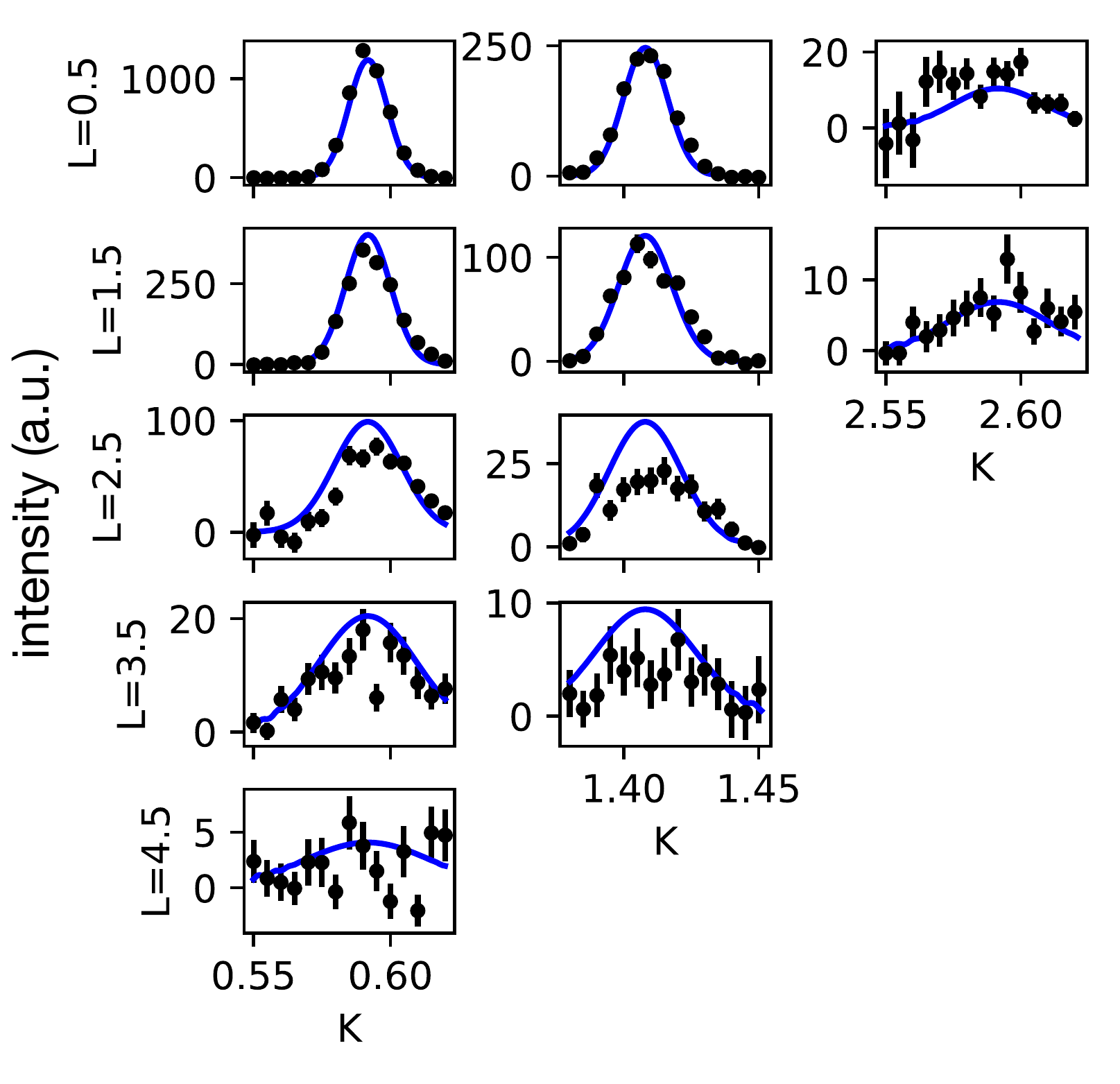}
\end{center}
\caption{Neutron scattering intensity (black points) of CrBr$_2$ at 3.7 K across various magnetic peaks, with the $L$ values labeled on the vertical axis. The blue lines show calculate intensity, assuming spins rotating in the $ac$ helical plane. For each data set, a background was subtracted consisting of the average of scans performed within $20 \leq T \leq 23$ K.}
\label{fig:CrBr2Inten}
\end{figure}

\begin{figure}[h]
\begin{center}
\includegraphics[width=8.6cm]
{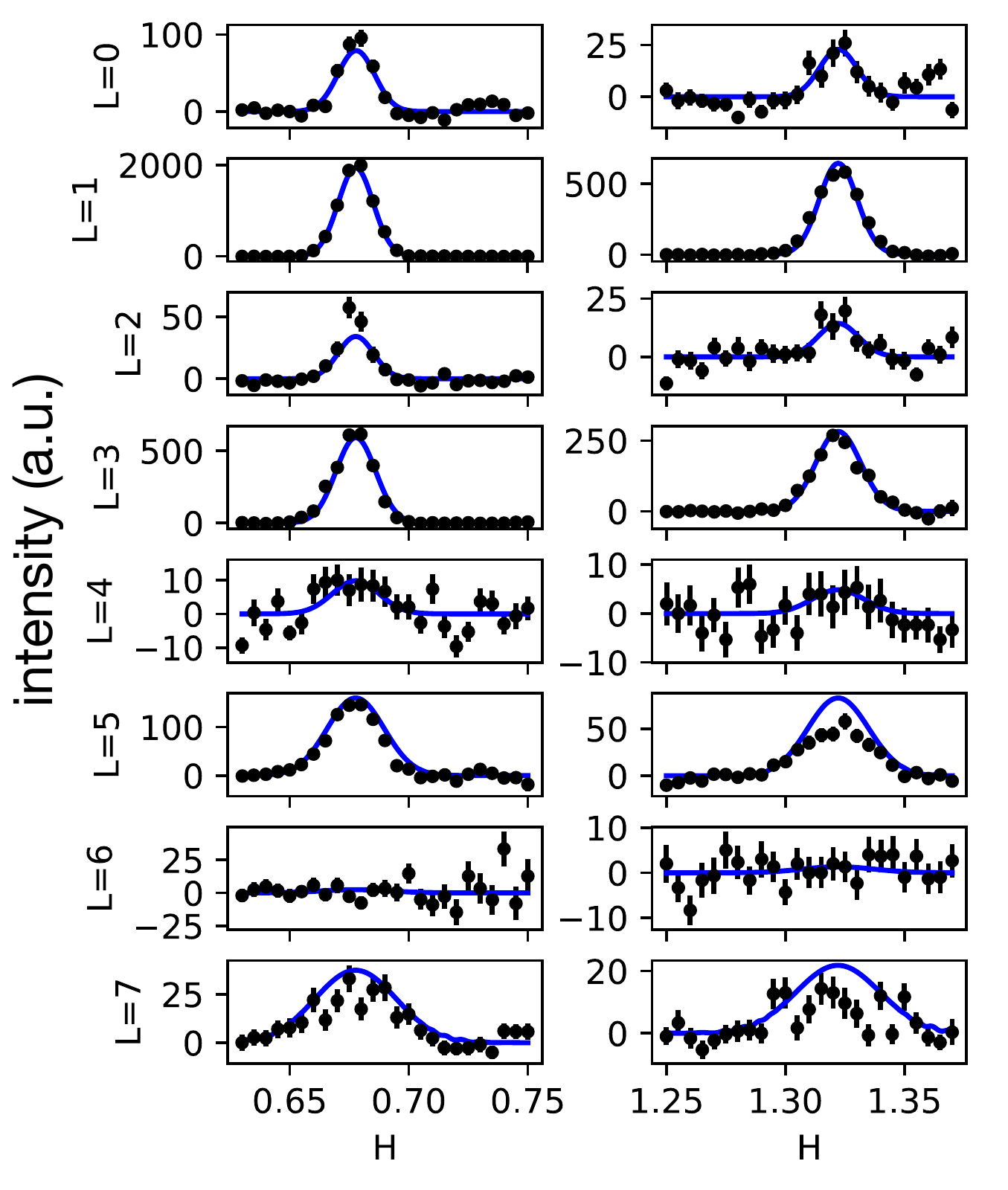}
\end{center}
\caption{Neutron scattering intensity (black points) of CrIBr at 3.8 K across various magnetic peaks, with the $L$ values labeled on the vertical axis. The blue lines show calculated intensity, assuming spins rotating in the $bc$ plane, corresponding to $\cos(\theta)(\psi_3 + i \psi_5) + i \sin(\theta)(\psi_6 + i \psi_2)$, with $\theta=10^{\circ}$. For each data set, a background was subtracted consisting of the average of scans performed within $13 \leq T \leq 14$ K.}
\label{fig:CrIBrInten}
\end{figure}

To determine the magnetic ordering of CrBr$_2$ and CrIBr, we performed elastic neutron scattering measurements on the VERITAS triple-axis spectrometer. Intensity was measured at one point in reciprocal space at a time. A typical scan consisted of a series of points across a Bragg peak, as shown by the arrow in Fig.\ \ref{fig:reciprocalSpaceMap}. Data for CrBr$_2$ and CrIBr are shown in Figures \ref{fig:CrBr2Inten} and \ref{fig:CrIBrInten}, respectively. The calculated intensity (the blue curves in these Figures) corresponds to the ideal integrated intensity for a magnetic peak (which depends on the model for the spin structure), convoluted with the resolution function, with an adjustable overall scale factor applied to all data taken for a given crystal and temperature; see Supplemental Materials for details \cite{supplement}. 

Fig.\ \ref{fig:reciprocalSpaceMap} illustrates the $0KL$ scattering plane of mono-CrBr$_2$ and the (essentially equivalent) $H0L$ scattering plane of ortho-CrIBr. Scans were done primarily in this scattering plane since it was the simplest plane in which the incommensurate magnetic peaks were present. 
However, intensity in this scattering plane is insensitive to position along the in-plane axis perpendicular to the ribbon chains, which makes it insensitive to the type of layer stacking present. 
For CrBr$_2$, additional measurements in the $H0L$ plane confirmed that the crystal consisted of a single twin with negligible diffuse scattering (see Supplemental Material \cite{supplement}.) For CrIBr, unfortunately, we were not able to perform measurements in the $0KL$ plane, and thus we have no direct knowledge of the layer stacking in this crystal (e.g., the presence of a monoclinic phase or of stacking disorder.) 
However, there are a couple of indications that our CrIBr crystal was mostly or entirely orthorhombic. First, only magnetic peaks for one value of the incommensurability $\delta$ were observed; assuming that the interlayer coupling has at least a small effect on $\delta$, we would expect peaks with multiple values of $\delta$ to be visible if a mono-CrIBr phase were present to a significant degree since the interlayer Cr-Cr bond distances change between mono-CrIBr and ortho-CrIBr. Second, peaks were seen at even as well as odd $L$, which implies orthorhombic stacking as we will discuss below. (There is also the possibility of disordered orthorhombic stacking, i.e., a sequence of layers with alternating orientation stacked according to an A/B as discussed in Section \ref{sec:SCXRD}. However, in a powder CrI$_2$ with nearly-random A/B stacking \cite{schneelochHelimagnetismCandidateFerroelectric2024}, the magnetic intensity was nonetheless consistent with a single value of $\delta$, and showed no clear evidence for disordered spin orientation despite the disordered layer stacking. Presumably, since the interlayer Cr-Cr bond distances are very nearly unchanged between A and B stacking types, the interlayer magnetic coupling would only be minimally affected by A/B stacking disorder.)

For CrBr$_2$, magnetic peaks were observed at $(0,K \pm \delta,L)$ for odd $K$ and half-integer $L$, with $\delta \approx 0.41$. 
No magnetic peaks were observed (at 3.7 K) along $(0K0)$ within $0.5 \leq K \leq 2.5$, or at $(0,2-\delta,0.5)$ or $(0,2+\delta,0.5)$. 
Such a pattern is consistent with a wavevector of $\mathbf{k} = (1,\delta,1/2)$, similar to CuCl$_2$ \cite{banksMagneticOrderingFrustrated2009} and CuBr$_2$ \cite{zhaoCuBr2NewMultiferroic2012}. 
The incommensurability corresponds to a $\sim$147$^{\circ}$ spin rotation between  consecutive Cr$^{2+}$ ions along the ribbon chains. If the $\sim$147$^{\circ} y$ helical twist were undone (where $y$ is the coordinate in lattice units along the ribbon chain), we would have AFM order within the layers and between neighboring layers along the $c$-axis. 

For CrIBr, essentially the same scattering plane was used, but labeled $H0L$ due to the use of the orthorhombic unit cell where the ribbon-chain axis is along $a$. 
The only sample peaks seen along $(H,0,1)$ for $0.5 \leq H \leq 2.6$ at 3.8 K were at $(1 \pm \delta, 0, 1)$ for $\delta \approx 0.322$, suggesting that peaks are only present at $H\pm\delta$ for odd $H$. 
Similar peaks were found at $(1\pm\delta,0,L)$ for both odd and even $L$ (equivalent to half- and whole-integer $L$ in the CrBr$_2$ coordinates.) This pattern is consistent with a propagation wavevector of $(\delta,0,0)$ corresponding to a helical angle of $\sim$116$^{\circ}$, in which the ``untwisted'' spins have intralayer AFM order, as for CrBr$_2$. Unlike CrBr$_2$, though, an arbitrary phase difference is allowed between the two layers in the orthorhombic unit cell, which explains the nonzero intensity at even $L$.

\begin{figure}[t]
\begin{center}
\includegraphics[width=8.6cm]
{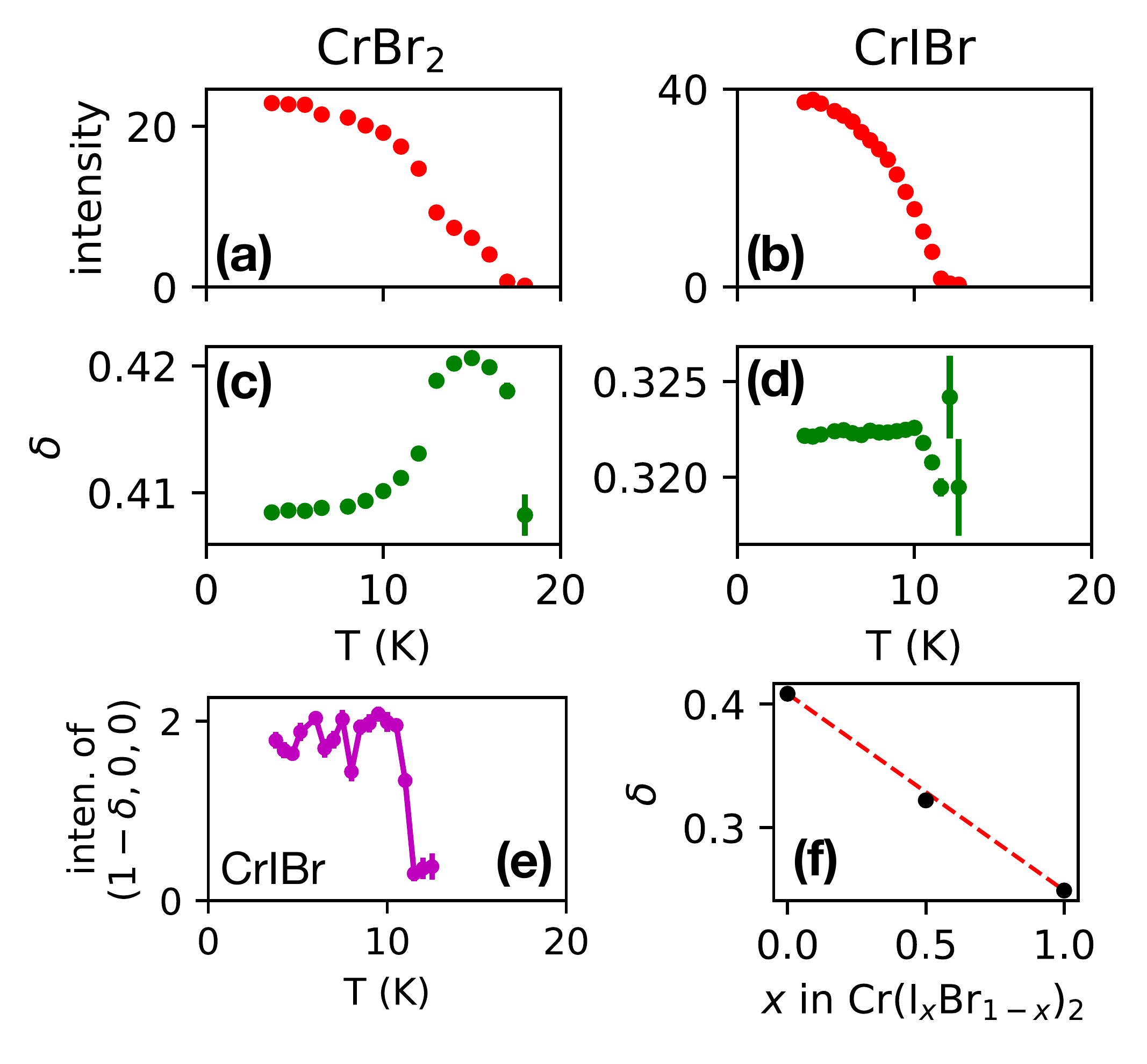}
\end{center}
\caption{(a-d) Integrated intensity (a,b) and the incommensurability $\delta$ (c,d) obtained from fits to scans across the magnetic Bragg peak at $(0,1-\delta,1/2)$ for CrBr$_2$ (a,c) and $(1-\delta,0,1)$ for CrIBr (b,d). (e) Intensity of the $(1-\delta,0,0)$ peak in CrIBr as a function of temperature. (f) Incommensurability vs.\ nominal iodine fraction. Values taken from the lowest-temperature points in (c) and (d) for CrBr$_2$ and CrIBr, and from the 8 K value for CrI$_2$ in Ref.\ \cite{schneelochHelimagnetismCandidateFerroelectric2024}. The red line connects the endpoints and is a guide to the eye. }
\label{fig:figTDep}
\end{figure}

Fig.\ \ref{fig:figTDep} shows parameters obtained from fits to the strongest magnetic peak at the equivalent locations of $(0,1-\delta,0.5)$ for CrBr$_2$ and $(1-\delta,0,1)$ for CrIBr. The integrated intensity and the incommensurability $\delta$ obtained from the  position are shown. From the intensity, we see that $T_N$ is about 17 K for CrBr$_2$ and 12 K for CrIBr, though a small remnant of a peak could be observed at slightly higher temperatures. 
The intensity for CrBr$_2$ shows a kink around 12 K, with an anomaly also present in $\delta$. 
The variation in $\delta$ over the measured temperature range is much larger, $\sim$0.01 reciprocal lattice units (r.l.u.), than the corresponding change of $\sim$0.002 r.l.u.\ in CrI$_2$. Additionally, in CrI$_2$, $\delta$ decreases on warming, while this parameter increases up to 15 K in CrBr$_2$. CrIBr shows a sharp change in $\delta$ at 10 K.
These unusual behaviors hint at a complexity that deserves further study.

In the Supplemental Material \cite{supplement}, we use representation analysis to list possible spin structures for CrBr$_2$ and CrIBr given their space groups and propagation vectors, then show the reduced $\chi^2$ values computed from an entire set of triple-axis scans for each spin structure.
For both materials, we found that a spin helix gave the best fit to the data, similar to CrI$_2$ \cite{schneelochHelimagnetismCandidateFerroelectric2024}, but in contrast to the cycloidal ordering reported for CuCl$_2$ \cite{banksMagneticOrderingFrustrated2009} and CuBr$_2$ \cite{zhaoCuBr2NewMultiferroic2012}. 
The spin-spiral rotation plane (i.e., whether the spin spiral is helimagnetic or cycloidal) is significant because, although there is interest in spin spiral magnets for potential multiferroic behavior, a ferroelectric polarization is only expected for \emph{cycloidal} magnetism \cite{tokuraMultiferroicsSpinOrigin2014}. However, for CuCl$_2$, the ferroelectric polarization can be suppressed with an applied magnetic field, which can change the spin-spiral rotation plane from cycloidal to helical \cite{sekiCupricChlorideTextCuCl2010}.  Presumably, the opposite effect (inducing a ferroelectric polarization with an applied magnetic field) may be possible for the chromium dihalides.

For the ortho-CrI$_2$ structure, an arbitrary phase difference is allowed by symmetry between the helices on the two layers of the unit cell, which we can characterize by an angle $\theta$. We define $\theta$ such that $\theta = 0^{\circ}$ corresponds to spins on opposite layers at $x=0$ being antialigned, where $x$ is the coordinate along the $a$-axis. Thus, the phase is accounted for by rotating spins on opposite layers by an alternating angle of $\pm \theta^{\circ}$ away from the $x=0$ anti-aligned configuration. If we assume our CrIBr crystal is purely orthorhombic, we find that, at 3.8 K, $\theta \approx 10.6(4)^{\circ}$ minimizes $\chi^2$, similar to the corresponding value of 15(1)$^{\circ}$ at 8 K for CrI$_2$ \cite{schneelochHelimagnetismCandidateFerroelectric2024}.

Interestingly, the intensity of the $(1-\delta,0,0)$ peak (Fig.\ \ref{fig:figTDep}(e)), has a markedly different temperature dependence from the $(1-\delta,0,1)$ peak (Fig.\ \ref{fig:figTDep}(b).) The $(1-\delta,0,0)$ intensity stays roughly constant until just before 11 K, at which point the intensity drops rapidly, while the $(1-\delta,0,1)$ intensity decreases continuously on warming. In the Supplemental Material \cite{supplement}, data from $L=0$ to 5 are shown, and we see that, for \emph{even} or \emph{odd} $L$, respectively, the temperature dependence resembles that of $(1-\delta,0,0)$ and $(1-\delta,0,1)$, respectively. Data are also presented for the $(1-\delta,0,0)$ and $(1-\delta,0,1)$ peaks of CrI$_2$ \cite{schneelochHelimagnetismCandidateFerroelectric2024}, in which the same pattern holds. This pattern indicates that the relative phase between alternating layers changes with temperature. For CrIBr, from 4 to 11.5 K, the $\theta$ value that minimizes $\chi^2$ increases from 10.6(4)$^{\circ}$ to 20(2)$^{\circ}$.

There is a striking linear dependence between the incommensurability $\delta$ (or the helical angle $\alpha = 2 \pi \delta$) and the iodine fraction $x$, as seen in Fig.\ \ref{fig:figTDep}(f). Further research is needed to see if this trend holds beyond these three points, and to obtain a full understanding of how the microscopic interactions change as a function of $x$ and temperature. Nevertheless, the data suggest the following compositional trends for the magnetic interactions. 
Assuming that the origin of the helimagnetism is intrachain frustrated magnetic interactions of the kind described by the $J_1$-$J_2$ model \cite{blundellMagnetismCondensedMatter2001}, one explanation is that electronic states tend to become delocalized as one substitutes an atom for its heavier congeners, in this case, Br with I. 
In the $J_1$-$J_2$ model \cite{blundellMagnetismCondensedMatter2001}, the helical angle $\alpha = 2 \pi \delta$ is given by $\cos(\alpha) = -J_1 / (4 J_2)$.  
Thus, if the magnitude of $J_2$ increases relative to $J_1$ due to a greater delocalization of electronic states in iodine relative to bromine, this would explain the observed movement of the helical angle closer to 90$^{\circ}$. 
(CrCl$_2$, with collinear AFM along its ribbon chains \cite{cableNeutronDiffractionStudies1960}, essentially has a helical angle of 180$^{\circ}$ and thus fits in with this trend.) 
An issue with this interpretation is that, in the $J_1$-$J_2$ model, $J_1$ would have to change from being weakly ferromagnetic (FM) for CrI$_2$ to sizably AFM for CrBr$_2$. (In detail, the helical angle of $89.7^{\circ}$ for CrI$_2$ implies $J_1 \approx -0.02 J_2$, while $\alpha=147^{\circ}$ for CrBr$_2$ implies $J_1 \approx 3.35 J_2$; $J_2$ must be AFM ($J_2 > 0$) for helimagnetism to occur in this model.) 

While $J_1$ may, indeed, be changing sign with composition, we argue that $J_1$ is likely AFM throughout the CrX$_2$ (X=Cl, Br, I) series, and longer-range intrachain interactions beyond the $J_1$-$J_2$ model may need to be accounted for to explain the changes in the helical angle. 
For CrCl$_2$, inelastic neutron scattering shows that the NN intrachain coupling is AFM (+1.13(13) meV) \cite{stoneQuasionedimensionalSpinWaves2013}. 
Given the structural similarity between the ribbon chains of CrCl$_2$, CrBr$_2$, and CrI$_2$, with bridging Cr-X-Cr angles of around 93.6$^{\circ}$ for CrCl$_2$ \cite{tracyCrCl2Paper,winkelmannStructuralMagneticCharacterization1997}, 91.6$^{\circ}$ for CrBr$_2$ \cite{tracyCrBr2Paper}, and within 90.8$^{\circ}$ to 91.8$^{\circ}$ for CrI$_2$ \cite{besrestStructureCristallineIodure1973,schneelochHelimagnetismCandidateFerroelectric2024,tracyCrI2MonoPaper}, it seems unlikely that the sign of the NN intrachain coupling would change. Since the NNN coupling is likely significant, as DFT calculations have suggested for CrI$_2$ \cite{schneelochHelimagnetismCandidateFerroelectric2024}, CuCl$_2$ \cite{banksMagneticOrderingFrustrated2009}, and CuBr$_2$ \cite{leeInvestigationSpinExchange2012}, the third-nearest neighbor coupling may also be significant. (DFT calculations \cite{schneelochHelimagnetismCandidateFerroelectric2024} suggest that intralayer interchain interactions may be too weak to explain the trend.)

As for the interlayer coupling, it is  notable that there is an apparent linearity between $\delta$ and iodine fraction despite an abrupt change in layer stacking somewhere between CrIBr and CrBr$_2$. If the interlayer magnetic coupling is substantial, as suggested for CrI$_2$ via DFT calculations \cite{schneelochHelimagnetismCandidateFerroelectric2024}, one would expect $\delta$ to have a significant deviation at some $x$ where the stacking change occurs. 
Of course, it is difficult to make firm conclusions about chemical trends from just three compositions, and the apparent linearity may be a coincidence, but the simplest explanation is that the interlayer coupling does not have a strong effect on the helical angle. 
While the interlayer coupling could certainly be weaker than calculated, it is also worth remembering that the contribution of each Cr-Cr bond in the Heisenberg model goes as $\cos(\Delta y)$, where $y$ is the displacement along the ribbon-chain axis in lattice units. Thus, Cr-Cr bonds that stretch further along the ribbon-chain axis, such as the intrachain NNN coupling with $|\Delta y| = 2$, would have a large impact on the helical angle, while the interlayer Cr-Cr bonds with $|\Delta y| = 0$ or $1/2$ would have zero or a small impact on the helical angle.

\subsection{Magnetic susceptibility of CrBr$_2$}

\begin{figure}[h!]
\begin{center}
\includegraphics[width=8.6cm]
{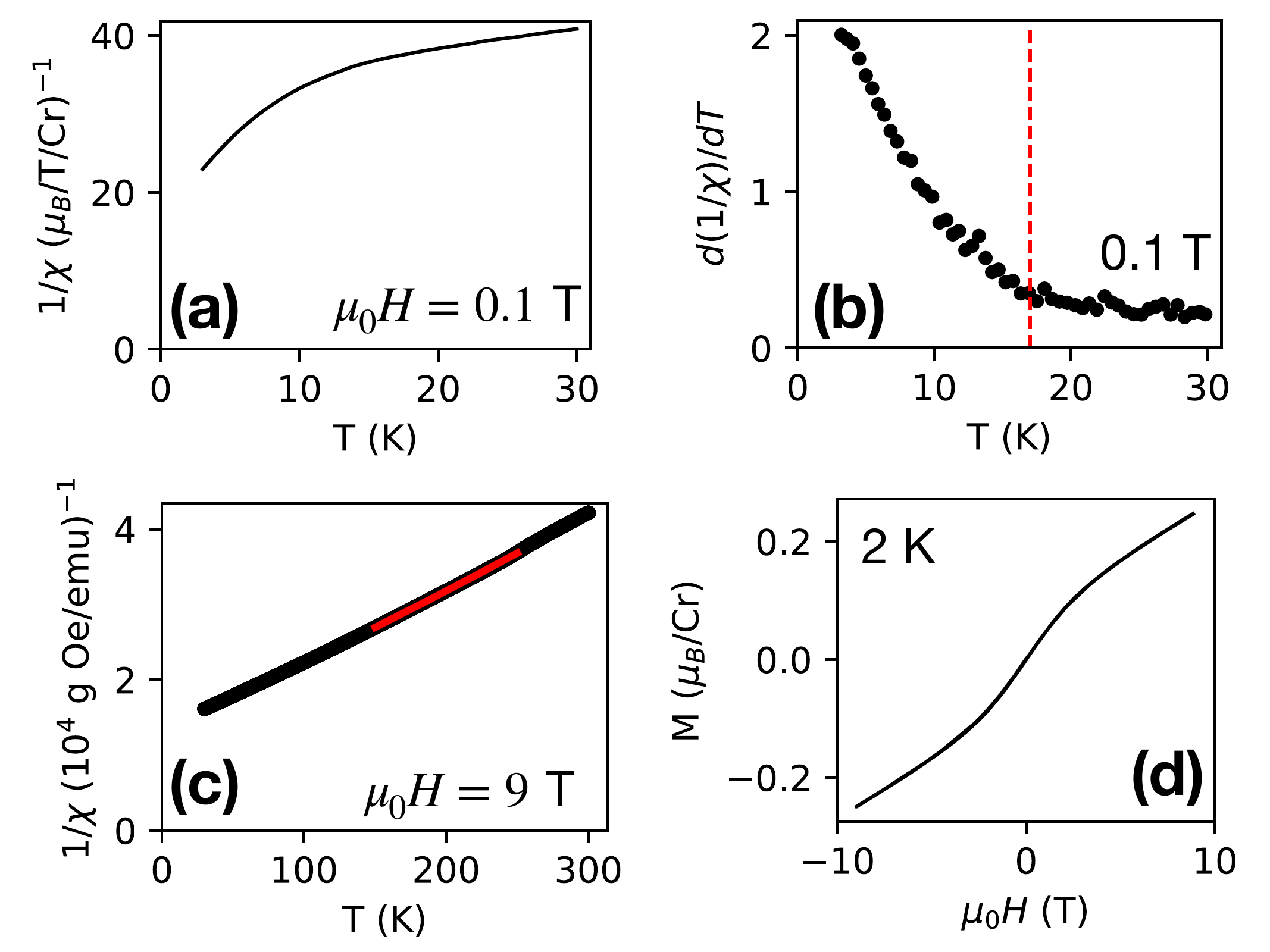}
\end{center}
\caption{Magnetization data taken on a CrBr$_2$ crystal with field applied in-plane. (a) Inverse susceptibility below 30 K at a field of $\mu_0 H = 0.1$ T, taken on warming from 2 K. (b) The slope $d(1/\chi)/dT$ of the data in (a), rebinned to improved statistics. The vertical dashed line indicates $T_N=17$ K. (c) Inverse susceptibility, taken on cooling from 300 K at a field of 9 T. The red line shows the region used for a Curie-Weiss analysis. (d) Magnetization versus applied field at 2 K, which was looped from $9$ T to $-9$ T and back, though no hysteresis was observed.}
\label{fig:susceptibility}
\end{figure}

In Fig.\ \ref{fig:susceptibility} we show magnetization data taken on a crystal of CrBr$_2$, with field applied in-plane. Duco cement was used to attach the crystal to a glass paddle sample holder. As for CrI$_2$ \cite{schneelochHelimagnetismCandidateFerroelectric2024}, the N\'{e}el transition in CrBr$_2$ can be seen as a very subtle signal in the magnetic susceptibility. Fig.\ \ref{fig:susceptibility}(a) shows the inverse susceptibility below 30 K, taken at a field of 0.1 T, and its slope with respect to temperature is shown in Fig.\ \ref{fig:susceptibility}(b). An upturn in $d(1/\chi)/dT$ can be seen below $\sim$17 K, consistent with the N\'{e}el temperature seen via neutrons. 

A Curie-Weiss analysis at much higher field (9 T) was done within $30 \leq T \leq 300$ K (Fig.\ \ref{fig:susceptibility}(c)) to verify the local-moment behavior of CrBr$_2$. A fit was done within $150 \leq T \leq 250$ K. (Data at higher temperatures were excluded due to a kink of unknown origin at $\sim$250 K.) It is common to include a temperature-independent term to a Curie-Weiss fit of the susceptibility (as was done for CuCl$_2$ \cite{banksMagneticOrderingFrustrated2009}), which incorporates possible contributions from the diamagnetism of the core electronic shells and a van Vleck paramagnetic term (as well as, presumably, the background contribution from the material used to secure the sample.) The functional form is, thus, $\chi = C/(T - T_{\textrm{CW}}) + \chi_0$, and the Curie constant $C$ can be converted to an effective magnetic moment $\sqrt{g^2 J (J+1)} \mu_B$. From the fit to the data in Fig.\ \ref{fig:susceptibility}(c), we obtained $\sqrt{g^2 J (J+1)} = 4.80(15) \mu_B$, $T_{\textrm{CW}} = -173(7)$ K, and $\chi_0 = -5(1) \cdot 10^{-6}$ emu/g/Oe, with the uncertainties estimated from the variation seen in fitting different temperature ranges. If we assume Cr$^{2+}$ has the high-spin configuration of $S=2$, $L=0$, and $g=2$, then we see that our effective moment is consistent with the ideal value of $\sqrt{g^2 J (J+1)} = 4.90 \mu_B$. 
The Curie-Weiss temperature of -173 K,  meanwhile, is much larger than the ordering temperature of 17 K, indicating the presence of magnetic frustration, unsurprisingly given that helimagnetism likely arises from competing AFM and FM interactions, as in the $J_1$-$J_2$ model. At 2 K, the magnetization vs.\ applied field behavior (Fig.\ \ref{fig:susceptibility}(d)) is non-hysteretic and approximately linear below $\mu_0 H \sim 2$ T.

\section{Discussion}
CrBr$_2$ joins a family of materials with 1) ribbon-chain structures formed due to Jahn-Teller distortion and 2) spin-spiral magnetic ordering that propagates along the ribbon chain direction. In this regards, most of the focus has been on Cu$^{2+}$ compounds such as CuCl$_2$ \cite{banksMagneticOrderingFrustrated2009} and CuBr$_2$ \cite{zhaoCuBr2NewMultiferroic2012}, but, as we've seen with CrI$_2$ \cite{schneelochHelimagnetismCandidateFerroelectric2024} and now CrBr$_2$, Cr$^{2+}$ compounds may have similar magnetic behavior. 
Our results suggest several open questions regarding the magnetic interactions in this system. 
First, while the tendency of electronic states to become more delocalized as atoms are replaced by those of heavier congeners explains the rough trend of the helical angle approaching 90$^{\circ}$ as we go from CrCl$_2$ to CrBr$_2$ to CrI$_2$, the fact that this angle is \emph{less than} 90$^{\circ}$ for CrI$_2$ implies, within the $J_1$-$J_2$ model, that $J_1$ is weakly FM, becoming sizably AFM for CrIBr and CrBr$_2$. 
Whether the NN intrachain coupling does, in fact, change sign with increasing Br fraction, or if additional interactions are needed to describe the magnetism, it is clear that more work needs to be done to understand the nature of the magnetic interactions. Also needing explanation is the degree to which interlayer magnetic coupling, suggested via calculations to be significant in CrI$_2$, contributes to the helimagnetism. 

Another open question is the source of the anomalous behavior in $\delta$ and the magentic peak intensity as a function of temperature. For both CrBr$_2$ and CrIBr, there is a change with temperature in the direction of movement of $\delta$, especially abrupt for CrIBr. These anomalies suggest a change in the relative magnitude of the interactions that determine $\delta$, but it seems implausible that structural changes in such a low temperature range could cause such behavior. Most likely, the effective magnetic interactions are modified with increasing spin disorder, but it is not clear how such changes could cause the abrupt changes observed. 

Finally, we note that it is reportedly possible to substitute Cr in CrI$_2$ with various transition metals, which provides an opportunity to study the effects of mixed spin configurations on the magnetism of CrX$_2$. For $M=$Mn, Fe, and Co, 1/3 substitution of $M$ for Cr in CrI$_2$ is reported to convert the ortho-CrI$_2$ stacking to the mono-CrI$_2$ stacking, while additional substitution (e.g., 2/3 substitution) results in CdCl$_2$-type or CdI$_2$ type structures where cooperative Jahn-Teller distortion (i.e., the ribbon chain) is no longer present \cite{guenSystemesCrI2MI2Ti1976,guenManganeseIIChrome1976,guenSystemeCrI2FeI2Diagramme1975}. (For Ni, the mono-CrI$_2$-type structure is reported at $\sim$50\% substitution rather than $\sim$33\%, and for V, a phase diagram could not be constructed.) Substituting Cr for other transition metals would certainly change the spin configuration on those atomic sites, which would modify the exchange interactions and possibly lead to new magnetic behavior.

\section{Conclusion}
Via elastic neutron scattering, we find that CrBr$_2$ and CrIBr have a similar helimagnetic spin structure as ortho-CrI$_2$: a spin helix that propagates along the ribbon chains. For CrBr$_2$, the helimagnetism sets in below $\sim$17 K with a propagation vector of about $\mathbf{k} = (1,0.41,0.5)$. For CrIBr, which tends to have the ortho-CrI$_2$ structure, the helimagnetism in CrIBr sets in below $\sim$12 K with a propagation vector of about $\mathbf{k} = (0.32,0,0)$. Both compounds have signs of intermediate magnetic transitions. 


\section*{Acknowledgments}
The work at the University of Virginia is supported by the Department of Energy, Grant number DE-FG02-01ER45927. This research used resources at the High Flux Isotope Reactor, a DOE Office of Science User Facility operated by Oak Ridge National Laboratory.

%

\clearpage

\section*{Supplemental Materials}
\beginsupplement

\subsection{Lack of diffuse scattering for CrBr$_2$ crystal measured by neutron scattering}

\begin{figure}[h]
\begin{center}
\includegraphics[width=8.6cm]
{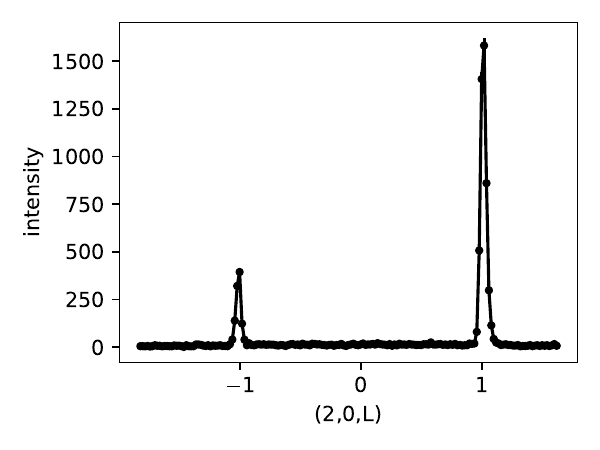}
\end{center}
\caption{Scan of elastic neutron scattering intensity along $(2,0,L)$. Peaks from only one twin can be observed (the other twin's peaks would be located at roughly $+0.23$ along $L$ from the first twin.)}
\label{fig:CrBr2NoDiffuse}
\end{figure}

In Fig.\ \ref{fig:CrBr2NoDiffuse}, we show neutron scattering intensity along the $(2,0,L)$ line of CrBr$_2$, which shows a lack of diffuse scattering that would indicate stacking disorder. We also see peaks from just one twin of CrBr$_2$; if the other twin were present, peaks at $(2,0,L+0.23)$ would be observed for integer $L$.

\subsection{Mathematical details for calculation of neutron scattering intensity}

The magnetic structure factor of a peak, given incommensurate ordering with a modulation vector $\mathbf{q}$, is \cite{kuindersmaMagneticStructuralInvestigations1981,carpenterElementsSlowNeutronScattering2015} 
\begin{multline}
\label{eq:magEq}
\mathbf{F}_M(\mathbf{G \pm q}) = \frac{\gamma_n r_0}{2 \mu_B}  \sum_j f(\mathbf{G} \pm \mathbf{q}) \frac{\mathbf{m}_j^{\mathbf{q}}}{2}  \times \\ 
\exp(i (\mathbf{G} \cdot \mathbf{d}_j \mp \phi_j)) 
\exp(-W_j).
\end{multline}
The $j$ index runs over every Cr$^{2+}$ ion in the unit cell.
Here, $\frac{\gamma_n r_0}{2 \mu_B} = 2.696$ fm, $f(\mathbf{Q})$ is the Cr$^{2+}$ magnetic form factor (assumed to be isotropic), $\mathbf{G}$ is the location of $(H,K,L)$, $\mathbf{m}_j^{\mathbf{q}}$ is the Fourier component of the $j$th magnetic moment in the unit cell ($\mathbf{m}_j^{\mathbf{q}}$ are in reference to the ``untwisted'' spins of the unit cell, before applying the helical rotation), $\mathbf{d}_j$ is the position of the $j$th ion, $\phi_j$ is a phase factor for the $j$th spin, and $W_j$ is the Debye-Waller factor. 
For a helix, $\mathbf{m}_j^{\mathbf{q}}$ would be of the form $\hat{u}_j + i \hat{v}_j$, where $\hat{u}_j$ and $\hat{v}_j$ are orthogonal and define the plane of helical rotation.

Convolutions to the resolution function for a triple-axis spectrometer were performed using the python package \textsc{neutronpy} \cite{fobes_neutronpy_2020}.

\subsection{Possible magnetic structures}
\begin{table}[h]
\begin{tabular}{ccc|cccccc}
  IR  &  BV  &  Atom & \multicolumn{6}{c}{BV components}\\
      &      &             &$m_{\|a}$ & $m_{\|b}$ & $m_{\|c}$ &$im_{\|a}$ & $im_{\|b}$ & $im_{\|c}$ \\
\hline
$\Gamma_{1}$ & $\psi_{1}$ &      1 &      0 &      2 &      0 &      0 &      0 &      0  \\
$\Gamma_{2}$ & $\psi_{2}$ &      1 &      2 &      0 &      0 &      0 &      0 &      0  \\
             & $\psi_{3}$ &      1 &      0 &      0 &      2 &      0 &      0 &      0  \\
\end{tabular}
\caption{Basis vectors for the IRs of CrBr$_2$, for group $C2/m$ at $\mathbf{k}=(0,1+\delta,0.5)$. This table was generated by \textsc{SARA\emph{h}}.}
\label{tab:basisVectorsCrBr2}
\end{table}

\begin{table}[h]
\begin{tabular}{c|c}
  CrBr$_2$ & $\chi^2_{\textrm{red}}$ \\
\hline
$\psi_1$  & 45 \\
$\psi_2$  & 9.5 \\
$\psi_3$  & 5.8 \\
$\psi_1$ + $i \psi_2$  & 20 \\
$\psi_1$ + $i \psi_3$  & 9.5 \\
\textcolor{blue}{$\psi_2$ + $i \psi_3$}  & \textcolor{blue}{3.9} \\
\end{tabular}
\caption{Reduced $\chi^2$ computed for the 4 K data shown in Fig.\ \ref{fig:CrBr2Inten} with respect to the calculated intensity of magnetic structures formed from the listed basis vectors. An overall scale factor was refined for each set to minimize $\chi^2$.}
\label{tab:chiSquaredCrBr2}
\end{table}

A systematic way to explore possible magnetic structures is to use representation theory analysis to list the irreducible representations (IRs) and the basis vectors associated with them, then find the basis vectors or combination thereof that best correspond to the data. 

Before continuing, we note that, in most of this paper, we use an ``AFM+twist'' picture of the magnetism for both CrBr$_2$ and CrIBr, in which we start with intralayer AFM order, then apply a twist corresponding to a modulation $\delta \approx 0.41$ or $0.32$, respectively. In this case, the magnetic Bragg peaks are located at $(H,K\pm\delta,L)$ for CrBr$_2$ and $(H\pm\delta,K,L)$ for CrIBr, with the requirement that $H+K$ is odd for both materials. (Of course, $L$ must be half-integer for CrBr$_2$ and integer for CrIBr.) For CrBr$_2$, a wavevector of $(1,\delta,0.5)$ captures this description, and is the form of the wavevector used to describe the similar helimagnetism of CuCl$_2$ \cite{banksMagneticOrderingFrustrated2009} and CuBr$_2$ \cite{zhaoCuBr2NewMultiferroic2012}. However, for representation analysis, it is simpler to describe the magnetism in an ``FM+twist'' picture, where we start with intralayer FM order, then apply a twist with modulation $1+\delta$. For example, in CrBr$_2$, the AFM+twist picture has the spins of an AFM structure rotated by $\sim$147$^{\circ} y$, where $y$ is the coordinate along the $b$-axis, while the FM+twist picture would have the spins of an FM structure rotated by $\sim$$(147+360)^{\circ} y$. For the FM+twist picture, the modulation vectors of CrBr$_2$ and CrIBr are $\mathbf{k} = (0,1+\delta,0.5)$ and $(1+\delta,0,0)$, and the Bragg peaks would be located at $(H,K\pm(1+\delta),L)$ and $(H \pm (1+\delta), K, L)$, respectively, for even $H+K$. These locations are equivalent to those given by the AFM+twist description, but, for convenience, we use the AFM+twist picture in most of this paper. For this section, though, we use the FM+twist description, and the Fourier components of the spins are equal within each layer.

We used \cite{willsNewProtocolDetermination2000} the software \textsc{SARA\emph{h}} to perform the representation theory analysis. 
For CrBr$_2$, Table \ref{tab:basisVectorsCrBr2} shows that there are two IRs: $\Gamma_1$, with the sole basis vector $\psi_1$, and $\Gamma_2$, with basis vectors $\psi_2$ and $\psi_3$. These basis vectors point along the crystallographic axes. Table \ref{tab:chiSquaredCrBr2} shows the reduced $\chi^2$ calculated from the 4 K data shown in Fig.\ \ref{fig:CrBr2Inten} for magnetic structures formed from various combinations of the basis vectors. We made the comparison with respect to the individual basis vectors, but also combinations of the form $\psi_j + i \psi_k$, which form helices for $j \neq k$. The form $\psi_2 + i \psi_3$ had the best agreement with data, as shown by both the $\chi^2_{\textrm{red}}$ value and the blue curves shown in Fig.\ \ref{fig:CrBr2Inten}. This form corresponds to a screw-like helix, i.e., with spins rotating in the $ac$ plane. A slight improvement of $\chi^2$ can be obtained by considering other structures, for example, ones in which the plane of the helix is tilted away from $ac$, but given the uncertainties associated with triple-axis spectrometer data, especially in a single scattering plane, a more precise exploration of the magnetic structure would require instrumentation more suitable for a magnetic structure refinement. 


\begin{table}[h]
\begin{tabular}{ccc|cccccc}
  IR  &  BV  &  Atom & \multicolumn{6}{c}{BV components}\\
      &      &             &$m_{\|a}$ & $m_{\|b}$ & $m_{\|c}$ &$im_{\|a}$ & $im_{\|b}$ & $im_{\|c}$ \\
\hline
$\Gamma_{1}$ & $\psi_{1}$ &      1 &      1 &      0 &      0 &      0 &      0 &      0  \\
             &              &      2 &     -1 &      0 &      0 &      0 &      0 &      0  \\
             & $\psi_{2}$ &      1 &      0 &      1 &      0 &      0 &      0 &      0  \\
             &              &      2 &      0 &      1 &      0 &      0 &      0 &      0  \\
             & $\psi_{3}$ &      1 &      0 &      0 &      1 &      0 &      0 &      0  \\
             &              &      2 &      0 &      0 &     -1 &      0 &      0 &      0  \\
$\Gamma_{2}$ & $\psi_{4}$ &      1 &      1 &      0 &      0 &      0 &      0 &      0  \\
             &              &      2 &      1 &      0 &      0 &      0 &      0 &      0  \\
             & $\psi_{5}$ &      1 &      0 &      1 &      0 &      0 &      0 &      0  \\
             &              &      2 &      0 &     -1 &      0 &      0 &      0 &      0  \\
             & $\psi_{6}$ &      1 &      0 &      0 &      1 &      0 &      0 &      0  \\
             &              &      2 &      0 &      0 &      1 &      0 &      0 &      0  \\
\end{tabular}
\caption{Basis vectors for the IRs for orthorhombic CrIBr, with space group $C m c 2_1$ and propagation vector 
$\mathbf{k}=(1+\delta,0,0)$, as generated by \textsc{SARA\emph{h}}. These basis vectors are identical to those considered for orthorhombic CrI$_2$ \cite{schneelochHelimagnetismCandidateFerroelectric2024}.}
\label{tab:basisVectorsCrIBr}
\end{table}

\begin{table}[h]
\begin{tabular}{c|c}
  CrIBr & $\chi^2_{\textrm{red}}$ \\
\hline
$\psi_1$  & 24 \\
$\psi_2$  & 37 \\
$\psi_3$  & 4.0 \\
$\psi_4$  & 38 \\
$\psi_5$  & 5.0 \\
$\psi_6$  & 37 \\
$\psi_1$ + $i \psi_2$  & 36 \\
$\psi_1$ + $i \psi_3$  & 5.0 \\
$\psi_1$ + $i \psi_5$  & 9.5 \\
$\psi_1$ + $i \psi_6$  & 36 \\
$\psi_2$ + $i \psi_3$  & 35 \\
$\psi_2$ + $i \psi_4$  & 37 \\
$\psi_2$ + $i \psi_6$  & 37 \\
$\psi_3$ + $i \psi_4$  & 29 \\
\textcolor{blue}{$\psi_3$ + $i \psi_5$}  & \textcolor{blue}{3.0} \\
$\psi_4$ + $i \psi_5$  & 24 \\
$\psi_4$ + $i \psi_6$  & 37 \\
$\psi_5$ + $i \psi_6$  & 32 \\
\end{tabular}
\caption{$\chi^2_{\textrm{red}}$ for the CrIBr data at 4 K (shown in Fig.\ \ref{fig:CrIBrInten}), computed with respect to the calculated intensity of magnetic structures formed from the listed basis vectors. An overall scale factor was refined for each set to minimize $\chi^2$.}
\label{tab:chiSquaredCrIBr}
\end{table}

For CrIBr, the analysis is very similar to that done for CrI$_2$ \cite{schneelochHelimagnetismCandidateFerroelectric2024}, though the data here (Fig.\ \ref{fig:CrIBrInten}) are a series of triple-axis spectrometer scans across certain Bragg peaks rather than powder diffraction data. Additionally, we do not know the true layer stacking of our CrIBr crystal (such as the extent to which it may have the monoclinic phase) due to lack of measurements in the $0KL$ scattering plane. Intensity in the orthorhombic $H0L$ plane is independent of position along the $b$-axis, but the differences in layer stacking may still have an effect due to differences in the relative spin orientation between the layers. For a monoclinic phase (either CrBr$_2$-like or CuCl$_2$/CuBr$_2$/CrI$_2$-like), we expect to have complete AFM orientation between the layers, but for the orthorhombic phase, there could, in principle, be an arbitrary phase between the layers, which would lead to nonzero magnetic peak intensity for even $L$. For CrI$_2$, a $\sim$15$^{\circ}$ deviation from interlayer AFM order was seen (i.e., every other layer had spins rotated by $\sim$15$^{\circ}$ in alternating directions) \cite{schneelochHelimagnetismCandidateFerroelectric2024}. 

For the orthorhombic structure (with space group $Cmc2_1$), with propagation vector $\mathbf{k} = (1+\delta,0,0)$, Table \ref{tab:basisVectorsCrIBr} shows that there are two IRs, each with three basis vectors. Each basis vector has six components, three each for the two atoms belonging to separate layers. (In the FM+twist picture, though there are two atoms per layer in the unit cell, they have equal spin components, so only one atom per layer is listed in Table \ref{tab:basisVectorsCrIBr}.) Each IR has basis vectors corresponding to the $a$, $b$, and $c$ directions, but the IRs differ in having some basis vectors have certain spin components on different layers pointing in opposite directions. In Table \ref{tab:chiSquaredCrIBr}, we list the reduced $\chi^2$ obtained from a comparison between the data in Fig.\ \ref{fig:CrIBrInten} and various combinations of the basis vectors, which includes the individual basis vectors, as well as those of the form $\psi_j + i \psi_k$ in which $\psi_j$ and $\psi_k$ are noncollinear. The $\psi_j + i \psi_k$ structures correspond to helices, though their direction varies, as well as whether helices on different layers rotate in the same or opposite directions. 

We see that, of the magnetic structures listed in Table \ref{tab:chiSquaredCrIBr}, $\psi_3 + i \psi_5$ has the lowest $\chi^2_{\textrm{red}}$ value, corresponding to screw-like helices rotating in the $bc$ plane, in the same direction between the two layers. However, this structure would result in zero intensity for the magnetic peaks at even $L$, due to the 180$^{\circ}$ phase difference between the layers. To be precise, the ``phase difference'' is the angular rotation between spins at $x=0$ on opposite layers, where $x$ is the coordinate along the $a$-axis. We can improve the agreement further if we allow an arbitrary phase difference, expressed as, e.g., $\cos(\theta) (\psi_3 + i \psi_5) + i \sin(\theta) (\psi_6 + i \psi_2)$, which is equivalent to multiplying the spin components on different layers by $e^{i \theta}$ and $e^{-i \theta}$. A phase of $\theta = 10.6(4)^{\circ}$ reduced $\chi^2_{\textrm{red}}$ from 3.03 to 2.06. This phase was used for the calculated intensity in Fig.\ \ref{fig:CrIBrInten}. This value is similar to the $\sim$$15^{\circ}$ value for CrI$_2$ \cite{schneelochHelimagnetismCandidateFerroelectric2024}, though, again, we do not know how much (if any) of our CrIBr sample is monoclinic, or how much stacking disorder is present.

\subsection{Temperature dependence of various Bragg peaks for CrIBr and CrI$_2$}

\begin{figure}[h]
\begin{center}
\includegraphics[width=8.6cm]
{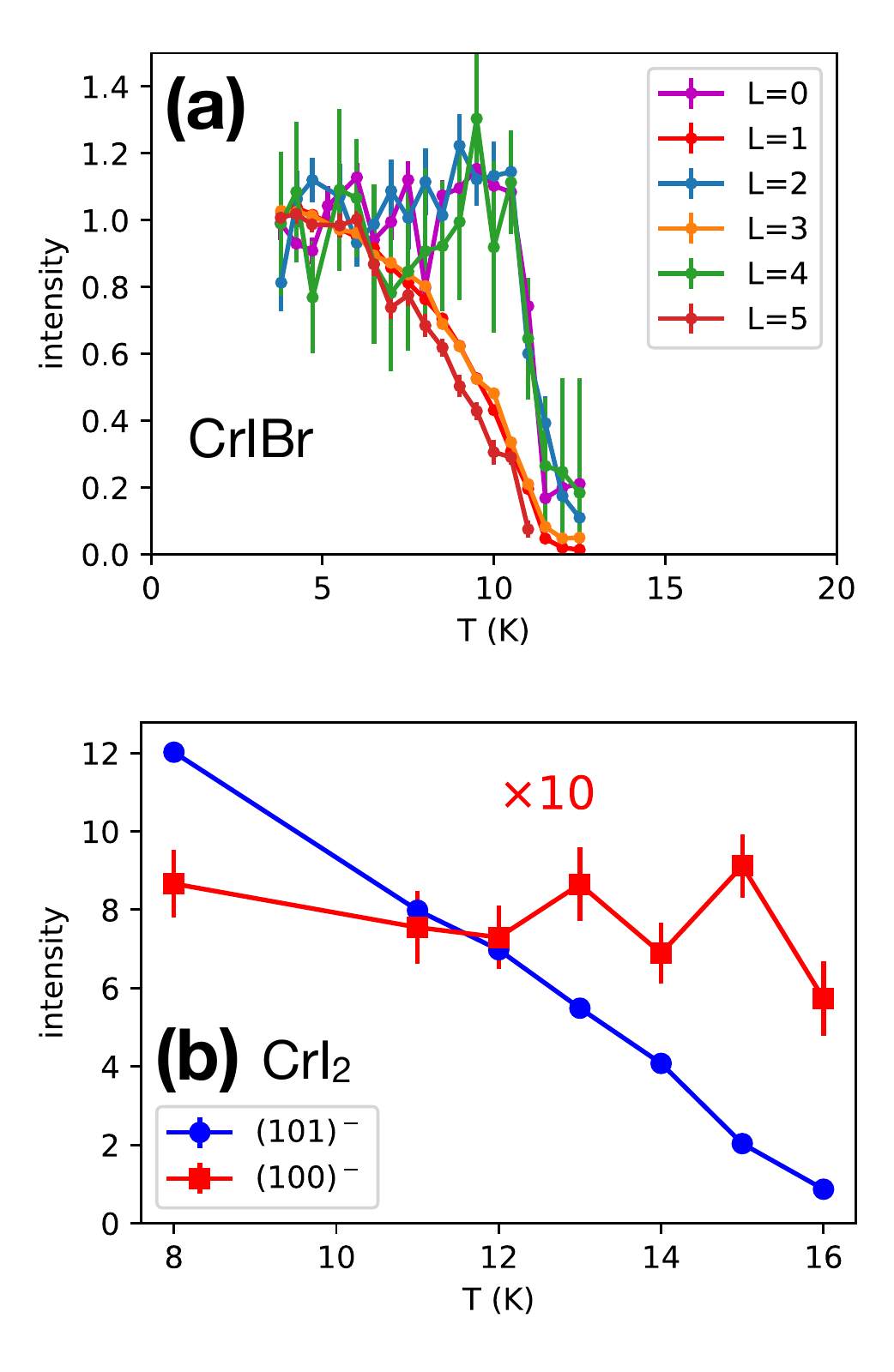}
\end{center}
\caption{Integrated intensity from fits of peaks at $(1-\delta,0,L)$ for (a) CrIBr, and (b) CrI$_2$ (b). The CrIBr data are the triple-axis spectrometry scans that are the focus of this work, normalized to their average intensity within $T \leq 6$ K. For CrI$_2$, the data fitted were the neutron diffraction data presented in Ref.\ \cite{schneelochHelimagnetismCandidateFerroelectric2024}, with $(1-\delta,0,1)$ and $(1-\delta,0,0)$ labeled as $(101)^-$ and $(100)^-$, respectively; the $(100)^-$ intensity is multiplied by 10. Both $(100)^-$ and $(101)^-$ are no longer visible by 17 K. The parameter $\delta$ was about 0.32 for CrIBr and 0.25 for CrI$_2$.}
\label{fig:AllLPeakTdep}
\end{figure}

In the main text, we discussed the fact that the temperature dependence of the CrIBr magnetic peak intensities differs depending on whether the peaks were located at $(1-\delta,0,L)$ for even or odd $L$. Here, in Fig.\ \ref{fig:AllLPeakTdep}(a), we show the intensity of the peaks at $(1-\delta,0,L)$ for $L$ ranging from 0 to 5. The intensity was the integrated intensity obtained from fits to triple-axis spectrometry scans, normalized to their low-temperature values. We see that the temperature dependence, indeed, splits into two groups, with odd-$L$ peaks decreasing steadily on warming, and even-$L$ peaks remaining constant until about 11 K. 

In Fig.\ \ref{fig:AllLPeakTdep}(b), we see that the same pattern holds for CrI$_2$ as for CrIBr. This Figure shows the fitted integrated intensity of the $(101)^-$ (at $(1-\delta,0,1)$) and $(100)^-$ (at $(1-\delta,0,0)$) magnetic Bragg peaks for CrI$_2$, extracted from the powder neutron diffraction data presented in Ref.\ \cite{schneelochHelimagnetismCandidateFerroelectric2024}. The $(101)^-$ peak intensity steadily decreases on warming, while $(100)^-$ remains roughly constant until 17 K, when both peaks are no longer clearly visible in the data.

\subsection{Temperature dependence of the overall scale factor and the interlayer $\theta$ angle in CrIBr}

\begin{figure}[h]
\begin{center}
\includegraphics[width=8.6cm]
{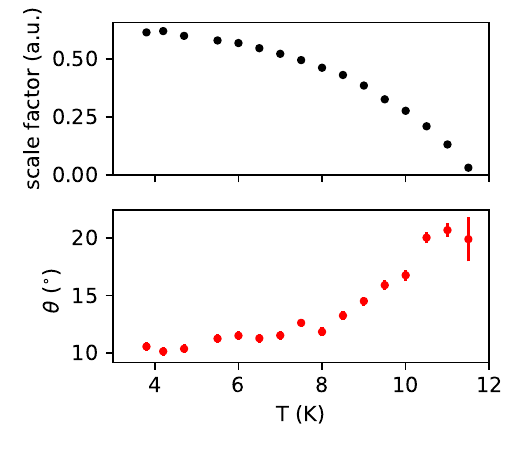}
\end{center}
\caption{Fitted parameters vs.\ temperature for CrIBr, specifically, the overall scale factor and the interlayer $\theta$ angle resulting from a spin structure model of the form $\cos(\theta) (\psi_3 + i \psi_5) + i \sin(\theta) (\psi_6 + i \psi_2)$.}
\label{fig:CrIBr_scaleFac_theta}
\end{figure}

Here, we show results from fitting the CrIBr data at various temperatures. We used the model described earlier, the $\cos(\theta) (\psi_3 + i \psi_5) + i \sin(\theta) (\psi_6 + i \psi_2)$ spin structure model, which describes screw-like helices with an angular difference $\theta$ between the layers (i.e., the deviation from AFM alignment between spins at $x=0$ on neighboring layers.) We see that $\theta$ increases from about 10$^{\circ}$ to 20$^{\circ}$ as temperature increases from 3.8 to 11.5 K, with most of the change occuring after 8 K.

\end{document}